\documentclass{article}

\usepackage{algorithm}
\usepackage{amsmath}
\usepackage{spconf,amsmath,graphicx}
\usepackage{caption}

\usepackage{booktabs}
\usepackage{mwe}
\usepackage{enumitem}
\usepackage{dingbat}
\usepackage{ragged2e} 
\usepackage{multirow}
\usepackage{booktabs,makecell, multirow, tabularx}

\title{Relate Auditory Speech to EEG by Shallow-Deep \\ Attention-based Network}
\name{Fan Cui\textsuperscript{1}, Liyong Guo\textsuperscript{1}, Lang He\textsuperscript{2}, Jiyao Liu\textsuperscript{3}, ErCheng Pei\textsuperscript{2}, Yujun Wang\textsuperscript{1}, Dongmei Jiang\textsuperscript{4}}

\address{\textsuperscript{1} Xiaomi Corp., Beijing, China \:\:  \textsuperscript{2} Xi'an University of Posts and Telecommunications, Xi’an, China  \\ \textsuperscript{3} Northwestern Polytechnical University, Xi’an, China \:\: \textsuperscript{4} Peng Cheng Laboratory, ShenZhen, China \\
\footnotesize{\texttt{\{cuifan, guoliyong, wangyujun\}@xiaomi.com, \:\:  \{langhe, ercheng.pei\}@xupt.edu.cn, jiangdm@nwpu.edu.cn}}}

\vspace{-0.5cm}
\begin{document}
%
\maketitle

\begin{abstract}

Electroencephalography (EEG) plays a vital role in detecting how brain responses to different stimulus. In this paper, we propose a novel Shallow-Deep Attention-based Network (SDANet) to classify the correct auditory stimulus evoking the EEG signal. It adopts the Attention-based Correlation Module (ACM) to discover the connection between auditory speech and EEG from global aspect, and the Shallow-Deep Similarity Classification Module (SDSCM) to decide the classification result via the embeddings learned from the shallow and deep layers.
Moreover, various training strategies and data augmentation are used to boost the model robustness.
Experiments are conducted on the dataset provided by Auditory EEG challenge (ICASSP Signal Processing Grand Challenge 2023). Results show that the proposed model has a significant gain over the baseline on the match-mismatch track.
\end{abstract}
\begin{keywords}
Electroencephalography, Attention
\end{keywords}
\vspace{-0.3cm}

\section{Introduction}

\label{sec:intro}
ICASSP 2023 Auditory EEG Challenge is designed to explore the relationship between auditory stimulus and evoked EEG signal. In this paper, we mainly focus on the first (match-mismatch) task. Traditional methods \cite{dmochowski2018extracting} adopt the linear model to fit the feature transform from the stimulus to the EEG signal. Recently, deep learning based methods have been proposed to improve the performance of relating speech with EEG signal\cite{monesi2020lstm, accou2021modeling}, where the feature transform modules are replaced by long context model, such as long short-term memory (LSTM), stacked dilated Convolutional blocks, etc.

In this study, we base our system design on the baseline\footnote{https://github.com/exporl/auditory-eeg-challenge-2023-code\label{baseline}}. Nevertheless, we combine attention structure in the proposed ACM module to extract correlation information from global view rather than employing stacked convolution layers to achieve a long context. Moreover, both shallow and deep layer embeddings are used to determine the similarity information in the SDSCM module. Additionally, we use random data augmentation and various training strategies to increase the robustness.

With the help of these efforts, the accuracy of our approach on the test dataset increases from 77\% to 80\%. The challenge's evaluation metrics for our method in the final blind testset are 78.94\%, 2\% higher than the baseline.

 
\vspace{-0.3cm}
\begin{figure}[t]
    \centering
    \includegraphics[width=\linewidth]{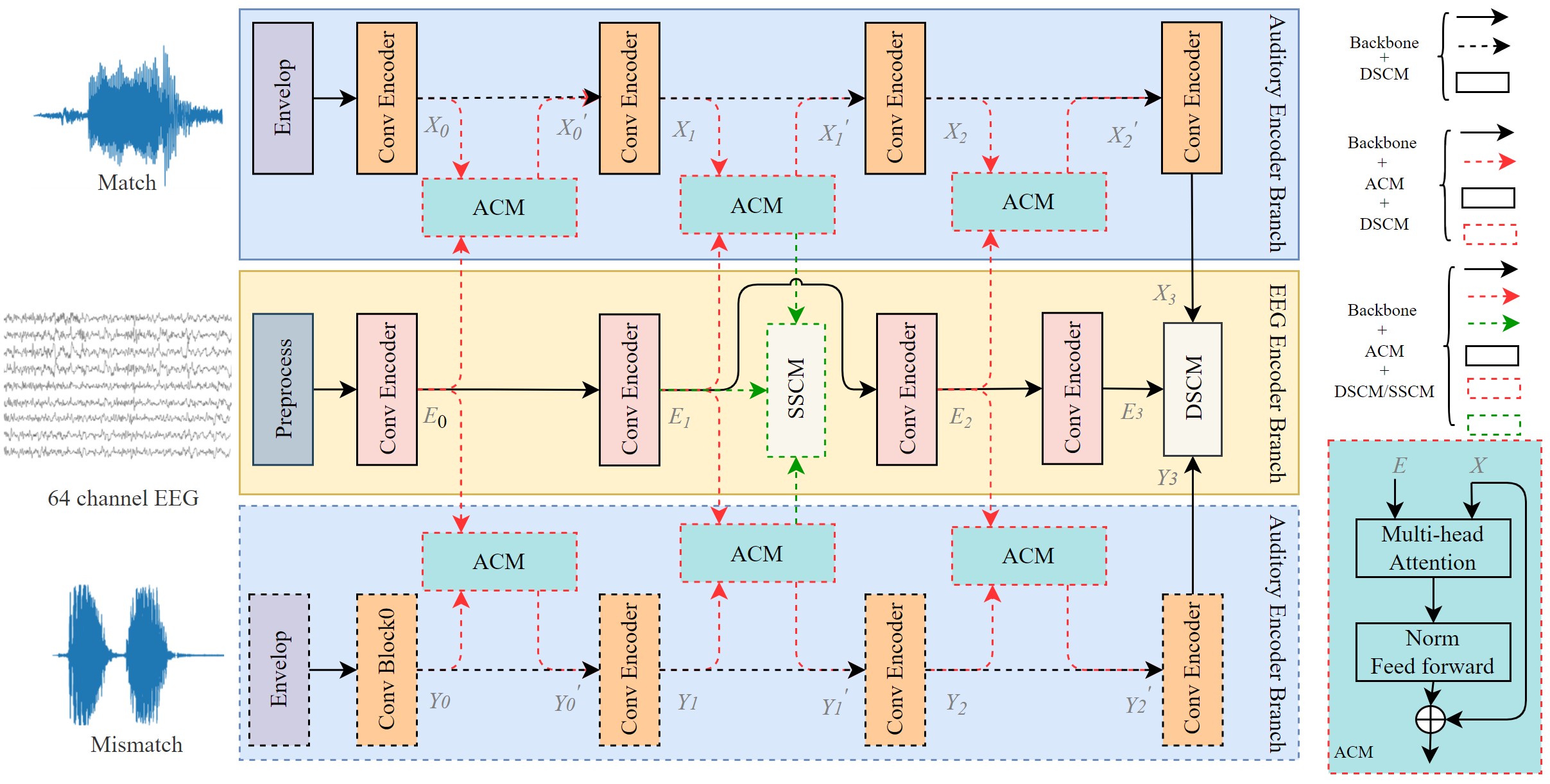}
    \caption{Illustration of the proposed model structures.}
    \label{fig:workflow}
    \vspace{-0.5cm}
\end{figure}

\section{Methodology}
\vspace{-0.3cm}
\subsection{Main Structure}

As shown in Figure~\ref{fig:workflow}, the system contains three inputs, including two audio stimuli and a slice of 64-channel EEG signal. The match and mismatch signals are fed to the auditory encoder branch with shared model parameters. The layers of convolution with kernel size 3 and powers of 2 dilation rate create a tree-like structure to enlarge the receptive field. Moreover, BatchNorm and ReLU layers are added to the outputs of convolutional layers.
Except for the four convolutional blocks, ACM blocks are introduced in the EEG encoder branch. The audio segments and EEG signals are converted to high-dimension representations using these encoder branches. Lastly, the match and mismatch signals are identified by the SDSCM blocks.
\vspace{-0.3cm}
\subsection{Attention-based Correlation Module}
\vspace{-0.1cm}
 Attention mechanism has been instrumental to make remarkable performance gains in many deep learning tasks.
Rather than processing the auditory and EEG branches separately, we adopt the ACM, consisting of a residual attention layer and a feed forward layer, to learn the relationship between EEG and auditory signal. The attention module maps query $X_{n}$ against key $E_{n}$ associated with candidate keywords, then presents re-weighted $X_{n}^{'}$ to emphasize the most important information.  
\vspace{-0.5cm}
\subsection{Shallow-Deep Similarity Classification Module}

As shown in Figure~\ref{fig:workflow},  SDSCM consists of Shallow Similarity Classifications Module(SSCM) and Deep Similarity Classification Module(DSCM).
The relation between the EEG and auditory signal can be evaluated at different scales. The extracted embedding has more specific information at the shallow layer, while the deep layer can gather profound semantic information. In this study, we compute the shallow-deep similarity embeddings using the encoder outputs as follows:
 \begin{eqnarray}
   E_{s}=Concat(\frac{X_1^{'}}{\left\|X_1^{'}\right\|_2}, \frac{E_1}{\left\|E_1\right\|_2}\cdot{\frac{Y_1^{'}}{\left\|Y_1^{'}\right\|_2}, \frac{E_1}{\left\|E_1\right\|_2})}  \\
   E_{d}=Concat(\frac{X_3^{'}}{\left\|X_3^{'}\right\|_2}, \frac{E_3}{\left\|E_3\right\|_2}\cdot{\frac{Y_3^{'}}{\left\|Y_3^{'}\right\|_2}, \frac{E_3}{\left\|E_3\right\|_2})}
\end{eqnarray}
where $X_1^{'},Y_1^{'},E_1,X_3^{'},Y_3^{'},E_3$ are the outputs of the convolutional blocks illustrated in Figure~\ref{fig:workflow}, $E_{s}$ and $E_{d}$ represent the similarity embeddings of the shallow and deep layers, respectively. In order to predict the binary classification probability, a Dense layer with one output is fed with the final classification embedding $E=Concat(E_{s}, E_{d})$.

\vspace{-0.3cm}

\section{EXPERIMENTS}
\label{sec:typestyle}
\vspace{-0.2cm}
\subsection{Data Processing} \label{data_proc}
\begin{figure}[h]
    \centering
    \includegraphics[width=\linewidth]{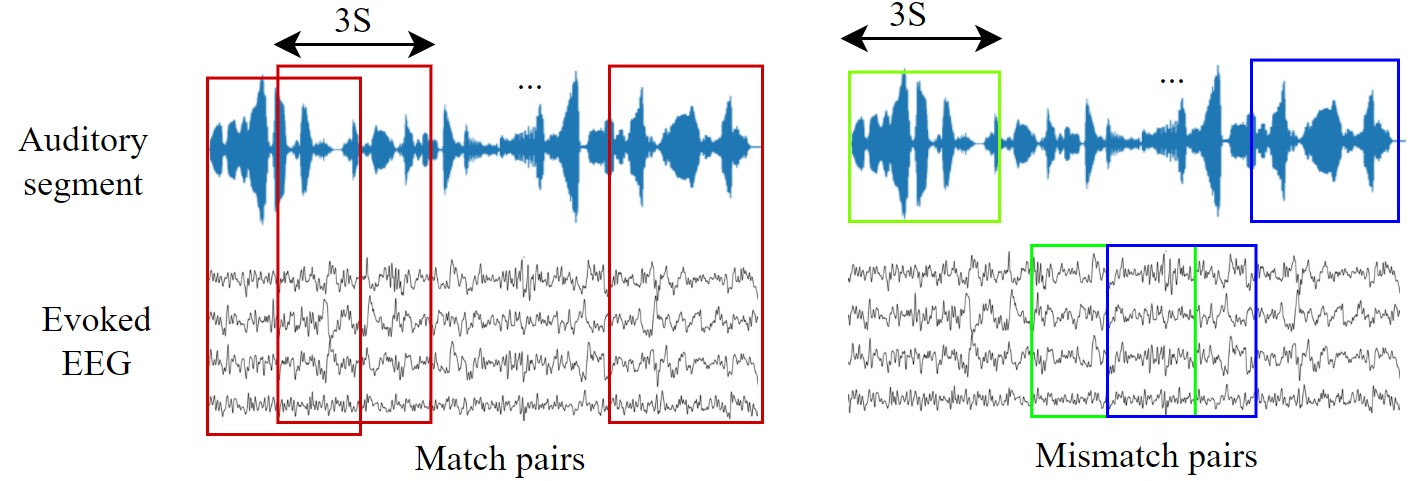}
    \caption{Illustration of training samplers generation.}
    \label{fig:traing_sampler}
    \vspace{-0.5cm}
\end{figure}

In the experiments, we use EEG dataset\cite{K3VSND_2023} provided by the EEG challenge, and split it into train-val-test subsets\textsuperscript{\ref{baseline}}. The generating training samplers are shown in Figure~\ref{fig:traing_sampler}. The matching pairs are produced using a sliding window of $window\_size=3s$ and $ window\_shift=Rand(1.0s,2.0s)$. The mismatch pairs are produced by randomly choosing the windows with an intersection of less than 35\% in order to produce various training data. According to the baseline\textsuperscript{\ref{baseline}}, the EEG signal is downsampled to 64 Hz, and the speech envelope is extracted for the inputs of the network. Besides, SpecAug\cite{park2019specaugment} is applied to the feature by masking and warping features channels and time steps. 
\vspace{-0.3cm}
\subsection{Training} \label{training}
\textbf{Training Configuration}: During training, Adam optimizer with learning rate 3e-4 is used, and when the validation loss stops dropping, we degrade the learning rate by a factor of 3. 64 samplers are created from 8 distinct subjects for each training batch. We also use weight decay (weight=0.0001) and dropout (drop\_rate=0.2) on each layer to prevent overfitting. The last ten models are utilized to do model averaging once the model has been trained for 100 epochs.

\vspace{-0.4cm}
\subsection{Results and Analysis} 
Table~\ref{tab:result1} displays accuracy comparison on test dataset. The proposed SDANet, as presented in Figure~\ref{fig:workflow}, contains four parts: Backbone, ACM, SSCM and DSCM. The backbone together with DSCM achieves 77.2\% accuracy. With our new data generation method, it achieves 2\% higher accuracy comparing official baseline approach\textsuperscript{\ref{baseline}}. After adding ACM, the accuracy further increases by 0.8\%. Finally, with both ACM and SDSCM modules, the accuracy is further improved to 80.2\%, which is 3.2\% above the baseline. On the final blind testset, our proposed model obtains a 78.94\% criteria determined in the official challenge description\textsuperscript{\ref{baseline}}.

\begin{table}[!ht]\footnotesize
    \vspace{-0.2cm}
\caption{Accuracy comparison results.}
    \vspace{-0.2cm}
\label{tab:result1}
\begin{tabular}{lllll|l}
\hline
\toprule
\multicolumn{5}{l|}{Methods}                                                                                                                                                                       & Acc    \\ \hline
\multicolumn{5}{l|}{Baseline}                                                                                                                                                                      & 77\%   \\ \hline
\multicolumn{1}{l|}{SDANet}                                                                          & \multicolumn{1}{l|}{Backbone} & \multicolumn{1}{l|}{DSCM} & \multicolumn{1}{l|}{ACM} & SSCM &        \\ \hline
\multicolumn{1}{l|}{\begin{tabular}[c]{@{}l@{}}Baseline\\ Data Generation\end{tabular}}              & \multicolumn{1}{l|}{\checkmark}        & \multicolumn{1}{l|}{\checkmark}    & \multicolumn{1}{l|}{}    &      & 77.2\% \\ \hline
\multicolumn{1}{l|}{\multirow{3}{*}{\begin{tabular}[c]{@{}l@{}}Our \\ Data Generation\end{tabular}}} & \multicolumn{1}{l|}{\checkmark}        & \multicolumn{1}{l|}{\checkmark}    & \multicolumn{1}{l|}{}    &      & 79\%   \\ \cline{2-6} 
\multicolumn{1}{l|}{}                                                                                & \multicolumn{1}{l|}{\checkmark}        & \multicolumn{1}{l|}{\checkmark}    & \multicolumn{1}{l|}{\checkmark}   &      & 79.8\% \\ \cline{2-6} 
\multicolumn{1}{l|}{}                                                                                & \multicolumn{1}{l|}{\checkmark}        & \multicolumn{1}{l|}{\checkmark}    & \multicolumn{1}{l|}{\checkmark}   & \checkmark    & 80.2\% \\ \hline
\end{tabular}
\end{table}

    \vspace{-0.3cm}
\section{CONCLUSION}
The objective of the study is to learn the relationship between auditory stimulus and EEG. We propose a SDANet model with ACM and SDSCM to increase the crossing linkages rather than treating them separately. 
Results show that our model improves the classification accuracy effectively.


\footnotesize
\bibliographystyle{IEEEbib}
\bibliography{mybib}

\end{document}